\documentclass[a4paper,11pt]{article}
\usepackage{pos}
\usepackage{slashed}

\newcommand{\ev}[1]{\left\langle #1 \right\rangle}
\newcommand{\evb}[1]{\big\langle #1 \big\rangle}

\newcommand{\Tr}{\text{Tr}}
\renewcommand{\Re}{\text{Re}}
\renewcommand{\Im}{\text{Im}}

\title{Equation of state of isospin asymmetric QCD with small baryon chemical potentials}
\ShortTitle{EoS of isospin asymmetric QCD with small baryon chemical potentials}

\author*[a]{Bastian B. Brandt}
\author[a,b]{Gergely Endr\H{o}di}
\author[b]{Gergely Mark\'o}

\affiliation[a]{Institute for Theoretical Physics, University of Bielefeld, D-33615 Bielefeld, Germany}

\affiliation[b]{Institute of Physics and Astronomy,
ELTE E\"otv\"os Lor\'and University,\\
P\'azm\'any P.\ s\'et\'any 1/A, H-1117 Budapest, Hungary}

\emailAdd{brandt@physik.uni-bielefeld.de}
\emailAdd{gergely.endrodi@ttk.elte.hu}
\emailAdd{marko@physik.uni-bielefeld.de}

\abstract{We extend our measurement of the equation of state of isospin asymmetric QCD to small baryon and strangeness chemical potentials, using the leading order Taylor expansion coefficients computed directly at non-zero isospin chemical potentials. Extrapolating the fully connected contributions to vanishing pion sources is particularly challenging, which we overcome by using information from isospin chemical potential derivatives evaluated numerically. Using the Taylor coefficients, we present, amongst others, first results for the equation of state along the electric charge chemical potential axis, which is potentially of relevance for the evolution of the early Universe at large lepton flavour asymmetries.}

\FullConference{The 41st International Symposium on Lattice Field Theory (LATTICE2024)\\
 28 July - 3 August 2024\\
Liverpool, UK\\}

\begin{document}
\maketitle

\section{Introduction}

One of the main ingredients for the phenomenological description of cosmological and astrophysical systems, as well as heavy-ion collisions, is the equation of state (EoS) of strongly interacting matter. While for most of the physical systems it is the baryon density which plays the major role, for some physical situations the charge density can actually be dominant. One example may be the evolution of the early Universe in the presence of large non-zero lepton flavour asymmetries~\cite{Wygas:2018otj,Middeldorf-Wygas:2020glx,Vovchenko:2020crk}. In any case, for a full description, knowledge about the EoS of strongly interacting matter in the full three-dimensional parameter space of light-quark chemical potentials is mandatory.

When the weak interactions can be neglected, individual quark densities are conserved and one can freely change the light-quark chemical potential basis. The most commonly used basis is the ``physical'' basis, where the individual quark chemical potentials are expressed through baryon ($B$), charge ($Q$) and strangeness ($S$) chemical potentials,
\begin{equation}
    \label{eq:phys-base}
    \mu_u=\frac{1}{3}\mu_B+\frac{2}{3}\mu_Q \,, \qquad \mu_d=\frac{1}{3}\mu_B-\frac{1}{3}\mu_Q \qquad \text{and} \quad \mu_s=\frac{1}{3}\mu_B-\frac{1}{3}\mu_Q-\mu_S \,.
\end{equation}
For simulations, a more convenient basis is the ``isospin'' basis, defined by
\begin{equation}
    \label{eq:iso-base}
    \mu_u=\mu_L+\mu_I \,, \qquad \mu_d=\mu_L-\mu_I \qquad \text{and} \quad \mu_s \,,
\end{equation}
where $\mu_L$ is the light-quark baryon chemical potential, $\mu_I$ the isospin chemical potential and one retains $\mu_s$ as simulation parameter. In this basis it is very easy to see when we run into a complex action problem. This is the case as soon as $\mu_L\neq0$ and/or $\mu_s\neq0$. For a pure isospin chemical potential, i.e. when $\mu_L=\mu_s=0$, the action is real and one can perform standard Monte-Carlo simulations to obtain the EoS~\cite{Son:2000xc,Kogut:2002tm,Kogut:2002zg}. In the past decade we have performed an extensive study of the properties of QCD at pure isospin chemical potential, including studies of the phase diagram at physical~\cite{Brandt:2017oyy,Brandt:2018omg,Brandt:2019hel,Cuteri:2021hiq} and smaller than physical~\cite{Brandt:2023kev} pion masses, as well as the EoS~\cite{Vovchenko:2020crk,Brandt:2017zck,Brandt:2018bwq,Brandt:2018wkp,Brandt:2021yhc,Brandt:2022fij,Brandt:2022hwy}, using improved actions.

For a full description of the physical systems mentioned above, it is mandatory to leave the pure isospin chemical potential axis. Since direct simulations are hampered by the complex action problem, the full parameter space can only be approached using indirect methods, such as the Taylor expansion method~\cite{Gottlieb:1988cq}. Up to now the Taylor expansion has been performed around the simulation points at vanishing chemical potential at temperature $T$. Here we will use simulation points at non-zero isospin chemical as novel expansion points (for a first account see~\cite{Brandt:2022fij}). The associated expansion is of particular importance for physical systems where the charge chemical potential plays the dominant role. A particularly interesting aspect of such systems is that for chemical potentials $\mu_I>m_\pi/2$ (equivalently, $\mu_Q>m_\pi$) one enters a phase with a Bose-Einstein condensate (BEC) of charged pions~\cite{Son:2000xc}. Due to the phase transition at the boundary of the BEC phase, standard Taylor expansions around $\mu_f=0$ cannot be used to learn about the EoS within the condensed phase.

\section{Taylor expansion from simulation points on the isospin axis}

The Taylor expansion from non-zero isospin chemical potential is an expansion in the light-quark baryon and strange quark chemical potentials and can be written as
\begin{equation}
    p(T,\mu_I,\mu_L,\mu_s)=p(T,\mu_I,0,0) + \sum_{n,m=1}^\infty
   \frac{1}{n!\,m!} \left. \frac{\partial^{n}\partial^{m}[p(T,\mu_I,\mu_L,\mu_s)]}
   {\partial \mu_L^n\,\partial \mu_s^m} \right|_{\mu_L,\mu_s=0} \big(\mu_L\big)^n\big(\mu_s\big)^m \,.
\end{equation}
In this proceedings article we are interested in the leading order of this Taylor expansion, i.e., the expansion to $O(\mu^2)$ in chemical potentials. The dominant Taylor coefficients at this order are the diagonal ones,
\begin{equation}
    \chi^L_2(T,\mu_I) \equiv \left. \frac{\partial^{2}[p(T,\mu_I,\mu_L,\mu_s)]} {\partial \mu_L^2} \right|_{\mu_L,\mu_s=0} \quad \text{and} \quad \chi^s_2(T,\mu_I) \equiv \left. \frac{\partial^{2}[p(T,\mu_I,\mu_L,\mu_s)]} {\partial \mu_s^2} \right|_{\mu_L,\mu_s=0} \,,
\end{equation}
which we need to extract together with the similarly defined mixed coefficient $\chi^{Ls}_{11}$ from our simulations at $\mu_I\neq0$.

In our studies we are using $N_f=2+1$ flavours of improved rooted staggered quarks with two levels of
stout smearing and physical quark masses (see Ref.~\cite{Brandt:2017oyy} for more details). The spontaneous symmetry breaking in the BEC phase necessitates the use of an explicit breaking term as a regulator, the pionic source term with parameter $\lambda$~\cite{Kogut:2002tm,Kogut:2002zg}. Simulations are done at $\lambda\neq0$ and physical results can be obtained after extrapolating to $\lambda=0$. This extrapolation is actually the main task for the analysis and has been facilitated in our previous studies through an improvement program, described in detail in Ref.~\cite{Brandt:2017oyy}. To be able to perform reliable $\lambda$-exrapolations for the Taylor expansion coefficients the first task is to adapt the improvement program to this more complicated observable. The improvement program consists of two parts, a valence quark improvement for the light quark observables and a leading order reweighting. We note that the reweighting does not depend on the observable, so that we can focus on the valence quark improvement in the following.

\subsection{Computation of Taylor expansion coefficients}

To adapt the valence quark improvement for the light-quark Taylor expansion coefficients, we first have to write the coefficients in terms of traces over inverses of the two-flavour fermion matrix for an isospin doublet, $\mathcal{M}$ and its derivatives, as given in Ref.~\cite{Brandt:2017oyy}, Eq.~(6). Second derivatives of the pressure can then be written as
\begin{equation}
    \label{eq:Tcoeffs-terms}
    \chi^{XY}_{11} = \frac{T}{V} \big[ \evb{c_{XY}} + \evb{c_X c_Y} - \evb{c_X} \evb{c_Y} \big] \,,
\end{equation}
where $X$ and $Y$ can be $L$ or $I$ and we have introduced the abbreviations
\begin{equation}
    c_X = \frac{1}{4}\Tr\Big[ \mathcal{M}^{-1} \frac{\partial \mathcal{M}}{\partial \mu_X} \Big]
    \quad\text{and}\quad
    c_{XY} = \frac{1}{4}\Tr\Big[ \mathcal{M}^{-1} \frac{\partial^2 \mathcal{M}}{\partial \mu_X\partial \mu_Y} \Big] - \frac{1}{4}\Tr\Big[ \mathcal{M}^{-1} \frac{\partial \mathcal{M}}{\partial \mu_X} \mathcal{M}^{-1} \frac{\partial \mathcal{M}}{\partial \mu_Y} \Big] \,,
\end{equation}
and the factors $1/4$ originate from rooting, i.e., are staggered specific. The first term on the right-hand side of Eq.~(\ref{eq:Tcoeffs-terms}) $c_{XY}$ is typically denoted as the connected part of the Taylor expansion coefficient and the other terms constitute the disconnected part.
In the following we will mostly focus on $\chi^{L}_{2}$ for simplicity. The improvement of $\chi^{Ls}_{11}$ and $\chi^{I}_{2}$ follow in complete analogy. $\chi^{s}_{2}$ does not include light quark contributions in the operator, so that no valence quark improvement is necessary.

For the purpose of the valence quark improvement as well as computational efficiency, one can reformulate the traces in terms of inverses of the matrix
\begin{equation}
    M=D^\dagger(\mu)D(\mu)+\lambda^2 \,,
\end{equation}
where $D$ is the massive one-flavour Dirac operator for the mass-degenerate isospin doublet. When reformulating the traces, additional factors appear in the fermion-matrix derivatives, so that we can write the individual traces eventually as
\begin{equation}
 \label{eq:Ttraces}
  \begin{array}{l}
    \displaystyle c_X = \hat{C}_X\Tr\Big[ M^{-1} \hat{O}_1 \Big]\,,\textrm{ with }\,\hat{O}_1 =  D^\dagger(\mu_I) \slashed{D}'(\mu_I)\,\textrm{ and} \vspace*{2mm} \\
    \displaystyle c_{XY} = \underbrace{\frac{1}{2}\Re\Tr\Big[ M^{-1} \hat{O}_{2} \Big]}_{\equiv c_{XY}^{(1)}} - \underbrace{\frac{1}{2}\Re\Tr\Big[ M^{-1} \hat{O}_1 M^{-1} \hat{O}_1 \Big]}_{\equiv c_{XY}^{(2)}}\,,\textrm{ with }\,\hat{O}_{2} = D^\dagger(\mu_I) \slashed{D}''(\mu_I)\,, 
  \end{array}
\end{equation}
and where we have introduced the operators $\hat{C}_L=(i/2)\,\Im$ and $\hat{C}_I=(1/2)\,\Re$, acting on the complex traces, the first and second derivatives of the derivative term of the Dirac operator, $\slashed{D}'$ and $\slashed{D}''$ and neglected terms of $O(\lambda^2)$, which will vanish in the $\lambda\to0$ limit.

\subsection{Valence quark improvement for the Taylor coefficients}
\label{sec:valtcoeff}

The representation of the traces in Eq.~(\ref{eq:Ttraces}) can be used to perform the valence quark improvement in terms of the singular values of the Dirac operator defined by
\begin{equation}
    D^\dagger(\mu_I)D(\mu_I)\varphi_n=\xi_n^2 \varphi_n \,.
\end{equation}
For the density type operators such as $c_X$ (note, that the isospin density $n_I\sim c_I$), the valence quark improvement proceeds by introducing an improvement term,
\begin{equation}
  \label{eq:impr-dens}
    \lim_{\lambda\to0}\ev{c_X}=\lim_{\lambda\to0}\ev{c_X-\delta^N_{c_X}} \quad\text{with}\quad \delta^N_{c_X} = \sum_{n=0}^{N-1} \hat{C}_X\big[ \varphi^\dagger_n \hat{O}_1 \varphi_n \big] \Big(\frac{1}{\xi_n^2+\lambda^2} - \frac{1}{\xi_n^2}\Big) \,,
\end{equation}
where $\delta^N_{c_X}$ is an approximation for the difference of the observable at vanishing and non-zero $\lambda$ using the lowest $N$ singular values to approximate the trace in Eq.~(\ref{eq:Ttraces}),
\begin{equation}
    \Tr\Big[ M^{-1} \hat{O}_1 \Big] = \sum_{n} \frac{\varphi^\dagger_n \hat{O}_1 \varphi_n}{\xi_n^2+\lambda^2} \approx \sum_{n=0}^{N-1} \frac{\varphi^\dagger_n \hat{O}_1 \varphi_n}{\xi_n^2+\lambda^2} \,.
\end{equation}
The efficiency of this valence quark improvement depends on the rate of convergence of the approximation with the number of singular values involved, $N$. Typical examples for the convergence of the improvement term for density type operators have been reported in Ref.~\cite{Brandt:2017oyy}.

For the disconnected terms of Eq.~(\ref{eq:Tcoeffs-terms}), the improvement term for the density type operators can be applied directly, leading to an improved extrapolation of the form\footnote{Note that there is also an alternative option of improving the first of the two disconnected terms by introducing a correction term directly for the squared trace. We have tested both versions and found that Eq.~(\ref{eq:discon-impro}) leads to a much stronger improvement for the $\lambda\to0$ limit.}
\begin{equation}
  \label{eq:discon-impro}
    \lim_{\lambda\to0} \Big[ \evb{c_X c_Y} - \evb{c_X} \evb{c_Y} \Big] = \lim_{\lambda\to0} \Big[ \evb{(c_X-\delta^N_{c_X})(c_Y-\delta^N_{c_Y})} - \evb{c_X-\delta^N_{c_X}} \evb{c_Y-\delta^N_{c_Y}} \Big] \,.
\end{equation}

To apply the method to the second of the connected terms $c_{XY}^{(2)}$, we have to obtain an improvement term for the fully connected summed two-point function. The trace of the fully connected two-point function contains two inverses of the matrix $M$, so that, for a full $\lambda$ improvement, we should insert two full sets of singular values. Truncating the two sums at $N$ singular values leads to the approximation
\begin{equation}
    \text{Tr}\Big[M^{-1} \hat{O}_1 M^{-1} \hat{O}_1 \Big] \approx \sum_{n,m=0}^{N-1} \frac{\varphi^\dagger_n \hat{O}_1 \varphi_m}{\xi_m^2+\lambda^2} \frac{\varphi^\dagger_m \hat{O}_1 \varphi_n}{\xi_n^2+\lambda^2} \,.
\end{equation}
With this approximation we can define the analogue of the improvement term from Eq.~(\ref{eq:impr-dens}) as
\begin{equation}
 \label{eq:lam-imp-susc}
 \delta^N_{c_{XY}^{(2)}} = \frac{1}{2} \sum_{n,m=0}^{N-1} \Re\big[ \varphi^\dagger_n \hat{O}_1 \varphi_m \varphi^\dagger_m \hat{O}_1 \varphi_n \big] \Big(\frac{1}{(\xi_n^2+\lambda^2)\,(\xi_m^2+\lambda^2)} - \frac{1}{\xi_n^2\,\xi_m^2} \Big) \,.
\end{equation}
The improvement for the term $c_{XY}^{(1)}$ proceeds as for the density type operators $c_X$, where the operator $\hat{O}_1$ is replaced by its second derivative analogue $\hat{O}_{2}$.

\begin{figure}[t]
 \vspace*{-2mm}
 \centering
 \includegraphics[width=.45\textwidth]{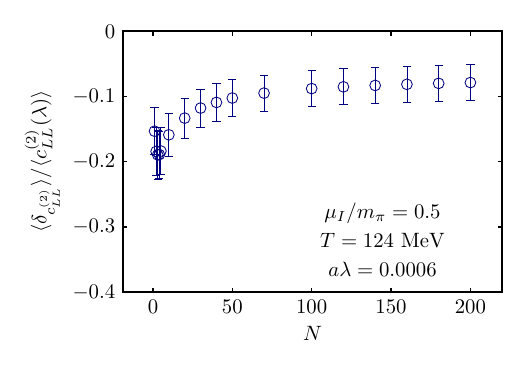}
 \includegraphics[width=.45\textwidth]{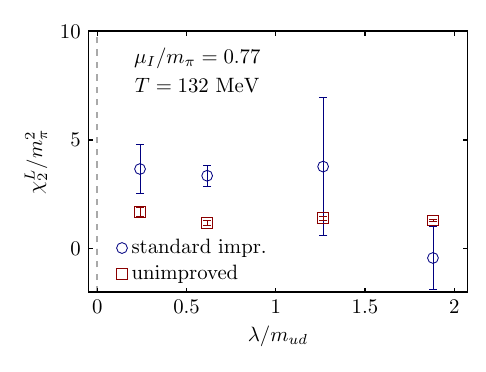}
 \vspace*{-3mm}
 \caption{{\bf Left:} Improvement term of the connected contribution $c_{LL}^{(2)}$ as obtained on a $24^3\times6$ lattice versus the number of included singular values normalized by $c_{LL}^{(2)}$ at the given $\lambda$.
 {\bf Right:} $\lambda$ dependence of the Taylor coefficient $\chi_2^L$ with and without improvement on a $24^3\times8$ lattice.}
 \label{fig:limpr}
\end{figure}

Outside of the BEC phase of condensed pions, the $\lambda$ extrapolations are well controlled and the improvement term has a comparably small effect. Within the BEC phase the contribution of the low singular values becomes more important, as can be seen from the improvement term of the connected contribution $c_{LL}^{(2)}$ which is plotted versus the number of singular values included in the left panel of Fig.~\ref{fig:limpr}. The plot shows that the effect of the improvement is large even if only one singular value is included and then converges to a slightly smaller value in magnitude when more singular values contribute. This behaviour is different compared to density type operators (see Fig.\ 5 of Ref.~\cite{Brandt:2017oyy}, for instance), where the full contribution develops from a sum over all singular values and the effect of the lowest singular value is not dominating the whole improvement term. Comparing Eqs.~(\ref{eq:impr-dens}) and~(\ref{eq:lam-imp-susc}) it is clear that this low mode dominance results from the squared inverses of the singular values in contrast to the linear dependence on the inverses for the density type operators. Unfortunately, the lowest singular value dominance also increases gauge fluctuations, drastically enhancing errors compared to density type operators. This can be seen for Taylor coefficients in the right panel of Fig.~\ref{fig:limpr}, where we show the $\lambda$-dependence of the improved (standard impr.) and unimproved observables obtained on a $24^3\times8$ lattice. The plot shows the increase in uncertainty, as well as the effect of the improvement, highlighting that it is beneficial for obtaining correct results in the $\lambda\to0$ limit.

\subsection{Density improvement for the computation of {\boldmath $\chi^L_2$}}
\label{sec:valtcoeff_impr}

\begin{figure}[t]
 \vspace*{-2mm}
 \centering
 \includegraphics[width=.45\textwidth]{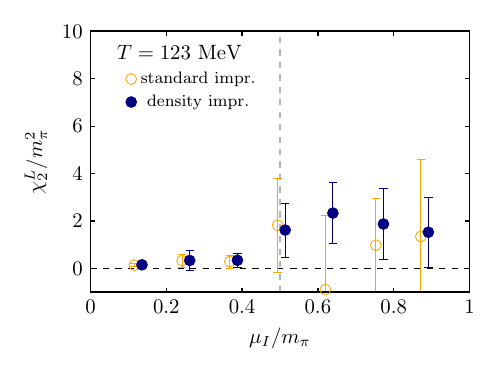}
 \includegraphics[width=.45\textwidth]{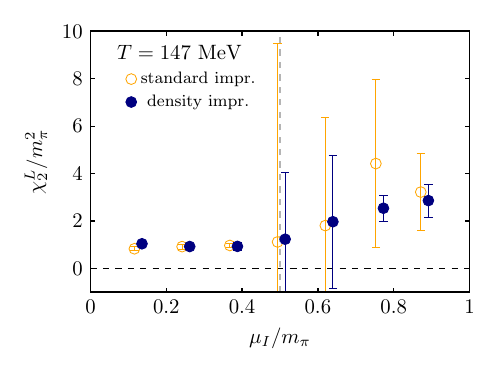}
 \vspace*{-3mm}
 \caption{Results for the Taylor expansion coefficient $\chi^L_2$ for two different temperatures obtained on $24^3\times8$ lattices from the usual $\lambda$ extrapolation (standard) using the standard version of the improved observables from Sec.~\ref{sec:valtcoeff} and the improvement version (improved) from Sec.~\ref{sec:valtcoeff_impr}. The results are slightly shifted to allow the comparison of the two sets of points.}
 \label{fig:tcoeff_lextra}
\end{figure}

The large uncertainties of the improved observables also lead to large uncertainties for the $\lambda=0$ extrapolated values in the BEC phase, as can be seen from the data labelled with ``standard impr.'' in the plots for the Taylor expansion coefficient $\chi_2^L$ of Fig.~\ref{fig:tcoeff_lextra}. We see that one cannot obtain a significant result within the BEC phase. Further improvement of the $\lambda$ extrapolations are needed.

We base our further improvement on the equality of the connected contributions of the second order Taylor expansion coefficients in $\mu_L$ and $\mu_I$ directions,
\begin{equation}
    c_{LL} = c_{II} \,,
\end{equation}
which follows directly from Eq.~(\ref{eq:Ttraces}). Using this relation, one can compute the coefficient $\chi_2^L$ via
\begin{equation}
    \chi^L_2(T,\mu_I) = \chi^I_2(T,\mu_I) + \frac{T}{V} \big[ \evb{(c_L)^2} - \evb{c_L}^2 - \big\{ \evb{(c_I)^2 - \evb{c_I}^2} \big\} \big] \,,
\end{equation}
where we have subtracted and added in the disconnected contributions of $\chi^I_2$ and $\chi^L_2$, respectively. The advantage is that $\chi^I_2$ can be computed directly at $\lambda=0$ using the improved results for $n_I$, together with a spline interpolation of its $\mu_I$-dependence~\cite{Brandt:2022hwy} to determine numerically $\chi_2^I=\frac{\partial n_I}{\partial \mu_I}$.

With this computation of $\chi_2^L$, what we denote as ``density improved'', only the disconnected contributions need to be extrapolated in $\lambda$, which reduces the uncertainties. The associated results are shown in Fig.~\ref{fig:tcoeff_lextra} with the label ``density impr.''. The plot indicates that uncertainties are indeed strongly reduced, so that significant results in the BEC phase can be obtained.

\section{The EoS at non-zero charge chemical potential}
\label{sec:muQeos}

\begin{figure}[t]
 \vspace*{-2mm}
 \centering
 \includegraphics[width=.45\textwidth]{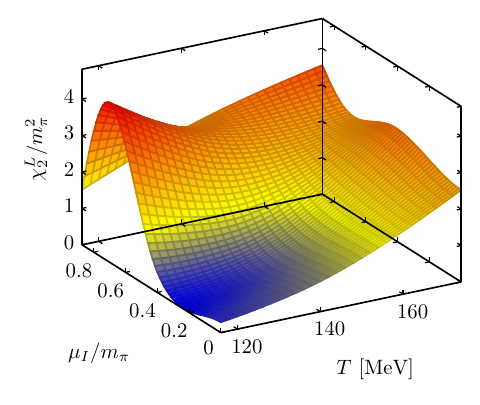}
 \includegraphics[width=.45\textwidth]{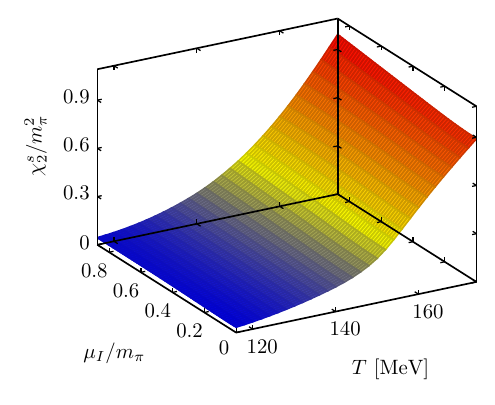} \\[-3mm]
 \includegraphics[width=.45\textwidth]{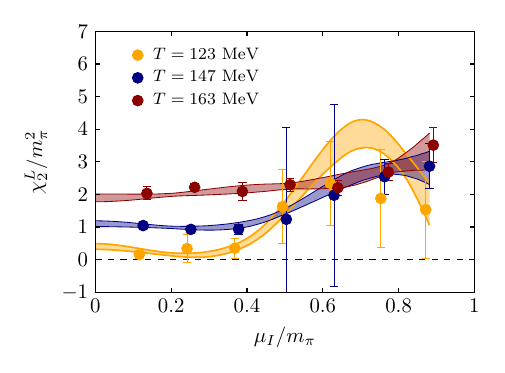}
 \includegraphics[width=.45\textwidth]{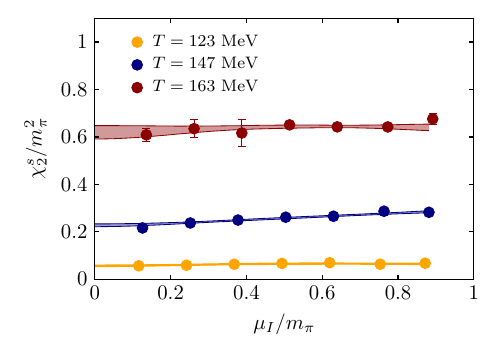}
 \vspace*{-3mm}
 \caption{Spline interpolation for the Taylor expansion coefficients $\chi^L_2$ (left) and $\chi^s_2$ (right). The top panel shows the full two-dimensional interpolation in the range of temperatures and chemical potentials where the EoS is also available from Ref.~\cite{Brandt:2022hwy} and the bottom panel shows the data together with the spline interpolation for three different temperatures.}
 \label{fig:tcoeff_interpol}
\end{figure}

As a first application for the Taylor expansion, we compute the EoS at pure charge chemical potential outside and within the BEC phase for the first time. This axis is of direct interest, since the early Universe in the presence of large lepton flavour asymmetries evolves in its vicinity~\cite{Wygas:2018otj,Middeldorf-Wygas:2020glx,Vovchenko:2020crk}. In our basis, a pure charge chemical potential $\mu_Q$ is obtained by setting
\begin{equation}
    \mu_I=\frac{\mu_Q}{2} \,, \quad \mu_L=\frac{\mu_Q}{6} \quad\text{and}\quad \mu_s=-\frac{\mu_Q}{3} \,.
\end{equation}
To compute the pressure on the $\mu_Q$ axis we first perform a two-dimensional spline interpolation of the Taylor expansion coefficients $\chi_2^L$, $\chi_2^s$ and $\chi_{11}^{Ls}$. $\chi_2^s$ is a simpler observable, since it is not a light-quark operator. The improvement of $\chi_{11}^{Ls}$ proceeds via the improvement of the isospin density and is also straightforward. As a boundary condition for the spline fits at $T=0$ we can make use of the fact that the Taylor coefficients vanish due to the Silver-Blaze property for any value of $\mu_I$. This is due to the fact that the first hadrons which can be excited, the neutron and the kaon, have non-vanishing masses in the full parameter space~\cite{Son:2000xc,Adhikari:2019mlf}. The results of the spline fits for the coefficients $\chi_2^L$ and $\chi_2^s$ are shown in Fig.~\ref{fig:tcoeff_interpol}. $\chi_{11}^{Ls}$ only gives a marginal contribution and remains of similar magnitude for all $\mu_I$.

\begin{figure}[t]
 \vspace*{-3mm}
 \centering
 \includegraphics[width=.44\textwidth]{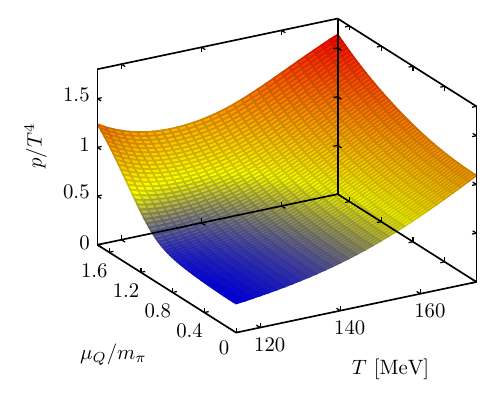}
 \includegraphics[width=.45\textwidth]{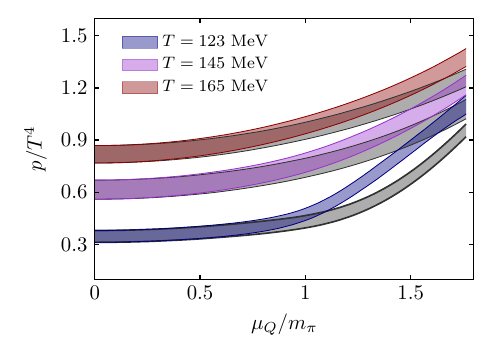}
 \vspace*{-4mm}
 \caption{Results for the pressure on the $\mu_Q$ axis obtained from leading order Taylor expansion starting from the $\mu_I$ axis. In the left panel we show the three-dimensional behaviour and in the right panel we show the results including uncertainties for three different temperatures. The light gray curves are the data for the pressure on the $2\mu_I/m_\pi$ axis from Ref.~\cite{Brandt:2022hwy} for comparison.}
 \label{fig:eos-muQ}
\end{figure}

Using the data for the Taylor expansion coefficients together with the data for the EoS at $\mu_I\neq0$ from Ref.~\cite{Brandt:2022hwy} we are now in the position to compute the EoS on the pure charge chemical potential axis. The results for the pressure are shown in Fig.~\ref{fig:eos-muQ}. In the right panel we compare the pressure to the one evaluated along the $\mu_I$ axis (gray curves). We see that the difference is most significant deep in the BEC phase and uncertainties only increase marginally in this interval.

\section{Conclusions}

In this proceedings article we have presented the details of the computations of the leading order Taylor expansion using simulation points on the pure isospin axis as novel expansion points. The key step in the analysis is the extrapolation of the Taylor expansion coefficients to vanishing regulator $\lambda$, facilitated by an improvement procedure. Part of the improvement is due to a light-quark operator (valence quark) improvement, which needs to be adapted to the case of $\chi^L_2$. We have discussed the improvement for the Taylor coefficients in detail and shown that the improvement, albeit necessary, introduces large uncertainties due to the fluctuations of the smallest singular values in the BEC phase. To obtain significant results in this regime we introduce and implement a method to reduce uncertainties by computing the connected part of the Taylor coefficient $\chi^L_2$ from $\chi^I_2$ directly at $\lambda=0$, see Sec.~\ref{sec:valtcoeff_impr}.

The final results for the Taylor coefficients are used to compute the EoS on the pure charge chemical potential axis for the first time in Sec.~\ref{sec:muQeos}. This equation of state is in the relevant region for the trajectories of the early Universe in the presence of large lepton flavour asymmetries~\cite{Wygas:2018otj,Middeldorf-Wygas:2020glx,Vovchenko:2020crk}. We note that the EoS on the $\mu_Q$ axis is obtained in Taylor expansion at leading order and is valid as long the expansion to this order is sufficient. In particular, it will break down once trying to expand through a phase boundary.

\acknowledgments
This work has been supported by the Deutsche Forschungsgemeinschaft (DFG, German Research Foundation) via CRC TRR 211 – project number 315477589, the Hungarian National Research, Development and Innovation Office (Research Grant Hungary 150241) and the European Research Council (Consolidator Grant 101125637 CoStaMM). The authors gratefully acknowledge the Gauss Centre for Supercomputing e.V. (\href{https://www.gauss-centre.eu}{\tt www.gauss-centre.eu}) for funding this project by providing computing time on the GCS Supercomputer SuperMUC-NG at Leibniz Supercomputing Centre (\href{https://www.lrz.de}{\tt www.lrz.de}) as well as enlightening discussions with Szabolcs Bors\'anyi, Attila P\'asztor and Lorenz von Smekal.

\providecommand{\href}[2]{#2}\begingroup\raggedright\endgroup


\end{document}